\documentclass{iaus}
\usepackage{graphicx}

\title[$^{12}$C/$^{13}$C ratio in stars] 
{$^{12}$C/$^{13}$C in atmospheres of red giants and peculiar stars}

\author[Ya.V.Pavlenko]   
{Yakiv V.Pavlenko$^1,^2$%
\affiliation{$^1$Main Astronomical Observatory of National Academy of
Sciences, 27 Zabolotnoho, Kyiv-127, 03680 Ukraine\break email: yp@mao.kiev.ua\\
$^2$Centre for Astrophysics Research, University of Hertfordshire,
College lane, Hatfield, AL10 9AB, UK\break email: yp@star.herts.ac.uk}}

\pubyear{2005}
\volume{228}  
\pagerange{1--7}
\date{?? and in revised form ??}
\jname{From Lithium to Uranium: Elemental Tracers of Early Cosmic Evolution}
\editors{V. Hill, P. Fran\c{c}ois \& F. Primas, eds.}
\begin{document}

\maketitle

\begin{abstract}
We determine the carbon isotopic ratios in
the atmospheres of some evolved stars in both globular clusters and the disk of
our Galaxy. Analysis of $^{12}$CO and $^{13}$CO bands at 2.3
micron was carried out using fits to observed spectra of red
giants and Sakurai's object (V4334 Sgr). The dependence of theoretical
spectra on the various input parameters was studied in detail. The computation of
model
atmospheres and a detailed abundance analysis was performed in a 
self-consistent fashion. A special procedure for
determining the best fits to observed spectra was used. We
show, that globular cluster giants with [Fe/H] $<$ -1.3 have a 
low $^{12}$C/$^{13}$C = 4 $\pm$ 1 abundance ration. In the spectra of 
Sakurai's object (V4334 Sgr) taken between 1997-98, the 2.3 micron 
spectral region is veiled
by hot dust emission. By fitting UKIRT spectra we
determined $^{12}$C/$^{13}$C = 4 $\pm$ 1 for the July, 1998 spectrum. CO
bands in the spectra of ultracool dwarfs are modelled as well.
\keywords{stars: carbon, chemically peculiar, abundances, AGB and
post-AGB, late-type, low-mass }
\end{abstract}

\firstsection 

\section{Introduction}

The carbon isotopes $^{12}C$ and  $^{13}C$ are produced inside low and
intermediate mass stars mainly. The carbon isotope ratio
$^{12}C$/$^{13}C$ is often used as a tracer of the nucleosynthetic
and chemical mixing processes inside stars. Fast evolving chemically
peculiar stars provide a unique tool for the verification of our
knowledge about the process of stellar evolution. In contrast,
some low mass dwarfs do not fuse $^1$H, these are known as brown dwarfs. 
These objects preserve their initial $^{12}C$ and $^{13}C$, and can be 
used for testing the Galaxy evolution theories.


\section{Procedure}

A  grid of model atmospheres, for red giants and Sakurai's object,
with the non-solar abundances was computed by using the model atmosphere
code SAM12(\cite{Pavlenko2003}). Molecular and atomic line opacities were
treated using the opacity sampling approach.
For ultracool dwarfs solar abundances are used. To compute their spectra
the  NextGen model atmospheres (\cite{nextg}) were used.

To determine the parameters of the ``best fit'', we use $F_{\nu}^x = \int
F^y_{\nu}G (x-y)dy $, where $F^y_{\nu}$ and $G(x-y)$ are
the synthetic fluxes and the broadening profile, respectively.
 We then find the minima of the 3D function
 $S(f_{\rm s}, f_{\rm h}, f_{\rm g}) =
   \sum \left ( 1 - f_{\rm h}  F^{\rm synt}/F^{\rm obs}  \right
   )^2   $,
where $f_{\rm s}$, $f_{\rm g}$, $f_{\rm g}$ are the wavelength
shift, normalisation factor, and half-width of $G(x-y)$,
respectively. The parameters $f_{\rm s}, f_{\rm h}$ and $f_{\rm
g}$ are determined by the minimisation procedure for every
computed spectrum (for details see \cite{jones},
\cite{PGE2004}.

\section{Sakurai's object(V4334 Sgr)}

V4334 Sgr is believed to be a low mass star undergoing a very late thermal
pulse (VLTP). After traversing the ``sholder'' of the post AGB
evolutionary track the star experiences a final He flash on the
way to white dwarf cooling track.

\cite{PG2002} found evidences that hot dust produces the
significant continuous emission at long wavelengths ($\lambda >$ 2
$\mu$m). When accounting for hot dust emission, our fits 
(\cite{PGE2004}) to the 2.3 $\mu$m CO bands yielded
 $^{12}$C/$^{13}$C   = 4 $\pm$ 1 for the July, 1998 spectra.This 
$^{12}$C/$^{13}$C ratio is consistent
with V4334 Sgr having undergone a VLTP.

\section{Red giants of globular clusters}
\cite{PJL2003} fit the synthetic spectra across 2.3 $\mu$m to
the observed spectra of a few globular cluster giants of different
metallicites. We found:
\begin{itemize}
\item that the M3 and M13 giants ([Fe/H]$<$ -1.3 dex) have about the same $^{12}$C/$^{13}$C,
\item we obtained some evidence of a lower $^{12}$C/$^{13}$C ratio in giants with lower [Fe/H],
\item giants of more metal rich clusters show larger dispersion of $^{12}$C/$^{13}$C,
\item more evolved stars show lower carbon abundance.
\end{itemize}

\section{Ultracool dwarfs}
\cite{PJ2003} compared observed and theoretical spectra of CO
bands of some M-dwarfs. In the spectra of oxygen-rich stars the CO bands are
severely blended by H$_2$O lines (see also
\cite{J2005}). We found the $^{13}$CO band at 2.375 $\mu$m
 is more useful for isotopic ratio determination
because the band suffers from far less contamination by water lines.
We then show the contamination by water bands appears to be stronger
for the $\Delta$v = 1 CO bands  at 4.5 $\mu$m.

\begin{acknowledgments}
I thank SOC of the IAU Symposium 228 for providing me
with an IAU travel grant. This work was partially
supported by visitor grants from the UK Particle 
Physics and Astronomy Research Council (PPARC), the UK Royal
Society, the Leverhulme Trust and SRG
from American Astronomical Society.  
Thanks to Greg Harris (UCL) for his help with the text.

\end{acknowledgments}


\begin{thebibliography}{}


\bibitem[Hauschildt at al. (1999)]{nextg}
{Hauschildt, P. H., Allard, F., Baron, E.} 1999, {\textit ApJ} 512, 377


\bibitem[Jones et al.(2002)]{jones} {Jones, H. R. A., Pavlenko, Ya., Viti, S.,
         Tennyson, J.} 2002, \textit{MNRAS} 330, 675



\bibitem[Jones, Pavlenko, Viti et al., (2005)]{J2005}
{Jones, H.R.A., Pavlenko, Ya., Viti, S., Barber, R.J.,
Yakovina, L.A., Pinfield, D., Tennyson, J.} 2005,
\textit{MNRAS} 358, 105



\bibitem[Pavlenko \& Geballe (2003)]{PG2002}
     {Pavlenko, Ya.V., Geballe, T. } 2002,
     \textit{Astron. Astrophys.} 390, 621

\bibitem[Pavlenko, Jones \& Longmore (2003)]{PJL2003}
     {Pavlenko, Ya.V., Jones, H.R.A., Longmore, } 2003,
     \textit{MNRAS} 345, 311

\bibitem[Pavlenko (2003)]{Pavlenko2003}
{Pavlenko, Ya,V.}
            {\textit Astron. Rept.} 2003, 47, 59

\bibitem[Pavlenko \& Jones (2003)]{PJ2003}
     {Pavlenko, Ya.V. \& Jones, H.R.A.} 2003,
     \textit{Astron. Astrophys} 397,967


\bibitem[Pavlenko, Geballe, Evans et al. (2004)]{PGE2004}
     {Pavlenko, Ya.V., Geballe, T.R., Evans, A.,  Smalley, B.,
Eyres, S.P.S.,Tyne, V.H., L.A. Yakovina, L.A.} 2004,
     \textit{Astron. Astrophys} 417, L39

\end{thebibliography}
\end{document}